# Max Born, Albert Einstein and Hermann Minkowski's Space-Time Formalism of Special Relativity

Galina Weinstein

This note is by no means a comprehensive study of Minkowski's space-time formalism of special relativity. The mathematician, Hermann Minkowski was Einstein's former mathematics professor at the Zürich Polytechnic. During his studies at the Polytechnic Einstein skipped Minkowski's classes. In 1904 Max Born arrived in the first time to Göttingen. Many years later Born wrote his recollections. In the summer of 1905, Minkowski and Hilbert led an advanced seminar on mathematical physics, on electrodynamical theory. Minkowski told Born later that it came to him as a great shock when Einstein published his paper in which the equivalence of the different local times of observers moving relative to each other was pronounced; for he had reached the same conclusions independently but did not publish them because he wished first to work out the mathematical structure in all its splendor. He never made a priority claim and always gave Einstein his full share in the great discovery. Indeed in his famous talk, "Space and Time" Minkowski wrote that the credit of first recognizing sharply that the time of the one electron is just as good as that of the other, i.e., that $t$ and $t'$ are to be treated the same, is of A. Einstein.

The mathematician, Hermann Minkowski was Einstein's former mathematics professor at the Zürich Polytechnic, and since 1902 a professor at Göttingen.

During his studies at the Polytechnic Einstein skipped Minkowski's classes. According to Einstein's biographer Carl Seelig, Minkowski's "lectures, at times badly prepared but full of creative power".[1] Einstein attended the following lectures of Minkowski: Geometry of Numbers, Function Theory, Potential Theory, Elliptic Functions, Analytical Mechanics, Variational Calculus, Algebra, Partial differential Equations, and Applications of Analytical Mechanics.[2]

Einstein could have gotten good mathematical training from his teachers at the Zürich Polytechnic. "At that time", says Anton Reiser, Einstein was less interested in mathematical speculation than in the visible process of physics".[3] Einstein felt that "the most fascinating subject at the time that I was a student was Maxwell's theory".[4] He found it difficult to accept for a long time the importance of abstract mathematics, and found high mathematics necessary only when developing his gravitation theory – he discovered the qualities of high mathematics around 1912.[5]

---

[1] Seelig Carl, *Albert Einstein: A documentary biography*, Translated to English by Mervyn Savill 1956, London: Staples Press, p. 27; Seelig Carl, *Albert Einstein; eine dokumentarische Biographie*, 1954, Zürich: Europa Verlag, p. 32.
[2] *The Collected Papers of Albert Einstein Vol. 1: The Early Years,1879–1902,* Stachel, John, Cassidy, David C., and Schulmann, Robert (eds.), Princeton: Princeton University Press, 1987, Doc. 28.
[3] Reiser, Anton *Albert Einstein: A Biographical Portrait*, 1930, New York: Dover, pp. 48-49.
[4] Einstein, Albert, "Autobiographical notes" in Schilpp, Paul Arthur (ed.), *Albert Einstein: Philosopher-Scientist*, 1949, La Salle, IL: Open Court, pp. 1–95; p. 31.

Jacob Johann Laub (an Austrian-born, who graduated under Professor Wilhelm Wien, professor of physics at the University of Würzburg) came to Bern to visit Einstein. Between 1908 and 1909 Einstein and Laub collaborated in three jointly works. In the opening paragraph of their first paper they wrote:[6]

"In a recent published study Mr. Minkowski has presented the fundamental equations for the electromagnetic processes in moving bodies. In view of the fact that this study makes rather great demands on the reader in its mathematical aspects, we do not consider it superfluous to derive here these important equations in an elementary way, which, is, by the way, essentially in agreement with that of Minkowski".

Einstein explained in his *Autobiographical Notes*,[7]

"I had excellent teachers (for example, [Adolf] Hurwitz, Minkowski), so that I should have been able to obtain a mathematical training in depth. I worked most of the time in physical laboratory, however, fascinated by the direct contact with experience. The balance of the time I used, in the main, in order to study at home the works of Kirchhoff, Helmholtz, Hertz, etc."

Einstein saw that it was split into numerous specialties – so he did not know what to do,[8]

---

[5] A girl named Barbara from Washington D.C. wrote Einstein on January 3, 1943, "I'm a little below average in mathematics", and she told Einstein that she had to work at it harder than most of her friends. Replying in English from Princeton on January 7, 1943, with his usual sense of humor, Einstein wrote in part as follows: "Do not worry about your difficulties in mathematics; I can assure you that mine are still greater". Calaprice, Alice (ed), *Dear Professor Einstein, Albert Einstein's Letters to And From Children*, 2002, New York: Prometheus Books, pp. 139-140.

[6] "In einer Kürzlich Vëroffentlichen Abhandlung hat Hr. Minkowski die Grundgleichungen für die elektromagnetischen Vorgänge in bewegten Körpern angegeben In Anbetracht des Umstandes, daß diese Arbeit in mathematischer Beziehung an der Leser Ziemlich große Anforderungen Stellt, halten wir es nicht für überflüssig, jene wichtigenGleichungen im folgenden auf elementaren Wege, der übrigens mit dem Minkowskischen im wesentlichen übereinstimmt, abzuleiten".
Einstein, Albert and Laub, Jacob, Johann, "Über die elektromagnetischen Grundgleichungen für bewegte Körper", *Annalen der Physik* 26, 1908, pp. 532-540; p. 532.
Arnold Sommerfeld's recollections of what Einstein said can further indicate his attitude towards mathematics before 1912: "Strangely enough no personal contacts resulted between his teacher of mathematics, Hermann Minkowski, and Einstein. When, later on, Minkowski built up the special theory of relativity into his 'world-geometry', Einstein said on one occasion: 'Since the mathematicians have invaded the theory of relativity, I do not understand it myself any more'. But soon thereafter, at the time of the conception of the general theory of relativity, he readily acknowledged the indispensability of the four-dimensional scheme of Minkowski". Sommerfeld, Arnold, "To Albert Einstein's Seventieth Birthday", in Einstein, 1949, in Schilpp, 1949, pp. 99-105; p. 102.
Abraham Pais also reported that before 1912 Einstein told V. Bergmann that he regarded the transcription of his theory into tensor form as "überflüssige Gelehrsamkeit" (superfluous learnedness). Pais, Abraham, *Subtle is the Lord. The Science and Life of Albert Einstein*, 1982, Oxford: Oxford University Press, p. 152.

[7] Einstein, 1949, in Schilpp, 1949, pp. 14-15.

[8] Einstein, 1949, in Schilpp, 1949, pp. 14-15.

"Consequently, I saw myself in the position of Buridan's ass, which was unable to decide upon any particular bundle of hay. Presumably this was because my intuition was not strong enough in the field of mathematics to differentiate clearly the fundamentally important, that which is really basic, from the rest of the more or less dispensable erudition. Also, my interest in the study of nature was no doubt stronger; and it was not clear to me as a young student that access to a more profound knowledge of the basic principles of physics depends on the most intricate mathematical methods. This dawned upon me only gradually after years of independent scientific work".

After 1912 Einstein regretted and wrote about his switch of attitude towards mathematics in the oft-quoted letter to Sommerfeld on October 29, 1912,[9]

"I am now occupied exclusively with the gravitational problem, and believe that I can overcome all difficulties with the help of a local mathematician friend [Marcel Grossmann]. But one thing is certain, never before in my life have I troubled myself over anything so much, and that I have gained great respect for mathematics, whose more subtle parts I considered until now, in my ignorance, as pure luxury! Compared with this problem, the original theory of relativity is childish".

Minkowski, who was the first to recognize the formal mathematical importance of Einstein's relativity theory, once admitted to his student, the physicist Max Born, "For me it came as a tremendous surprise, for in his student days Einstein had been a real lazybones. He never bothered about mathematics at all".[10]

Constance Reid, a popular American biographer wrote that Minkowski later often told his students in Göttingen, "Einstein's presentation of his deep theory [of relativity] is mathematically awkward – I can say that because he got his mathematical education in Zürich from me".[11] Reid did not give reference to this quotation and did not say from whom she heard this. It was probably just another of the "Einstein anecdotes" widespread in Göttingen. Reid wrote her book in English, and she visited in Göttingen; suppose that she heard this anecdote in Göttingen, then she had somehow to translate the quotation from German to English.

In 1907-1908 Minkowski developed the mathematical formalism for Einstein's relativity theory. He was occupied with the problem of electrodynamics theory at least from 1905.

In 1904 Max Born arrived in the first time to Göttingen. Many years later Born wrote his recollections. Born said that his stepmother gave him a letter of introduction to Minkowski, whom she had met years before at dancing lessons and balls at Königsberg.[12] After delivering the letter, Born received an invitation to dinner (German style, at 1:30 p. m.) from Frau Minkowski for Sunday. He was received most cordially and enjoyed an excellent meal. After the dinner Minkowski asked Born whether he would like to join in a little excursion to a ruined castle in the neighborhood, the 'Plesse'. They walked through the fields and woods to the 'Plesse'.[13] By August 1904 Born was David Hilbert's 'private assistant' – Hilbert offered Born this unpaid position at the end of his first year in Göttingen.[14]

---

[9] Einstein to Sommerfeld, October 29, 1912, *The Collected Papers of Albert Einstein (CPAE, Vol. 5): The Swiss Years: Correspondence, 1902–1914*, Klein, Martin J., Kox, A.J., and Schulmann, Robert (eds.), Princeton: Princeton University Press, 1993, Doc. 421.
[10] "Denn früher war Einstein ein richtiger Faulpelz. Um die Mathematik hat er sich überhaupt nicht gekümmert". Seelig, 1956, p. 28; Seelig, 1954, p. 33.
[11] Reid, Constance, *Hilbert*, 1970/1996, New-York: Springer-Verlag, p. 112.

Born recounted that Minkowski and David Hilbert were intimate friends from their school-days at Königsberg, and inseparable at the University. The dominant figure in mathematics in Germany and Göttingen at that time was Feix Klein. He had an immense influence on all things concerned with mathematics. He persuaded his faculty in 1895 to nominate Hilbert to a second professorship, to which the Prussian Ministry agreed. From that time on Hilbert was the star of mathematical Göttingen. But he missed his friend Minkowski and worked incessantly to get him called to Göttingen. In 1902 he succeeded; and a third professorship was offered to Minkowski.[15]

In the summer of 1905, Minkowski and Hilbert led an advanced seminar on mathematical physics, on electrodynamical theory. Participants of the seminar, Max Laue, Max Born, Max Abraham, Arnold Sommerfeld and others, studied the papers of Hendrik Antoon Lorentz, Henri Poincaré and others on the difficulties which the theories of the electrodynamics had run into as a result of Michelson's celebrated experiment. Born says that at that time the ether was considered well established, and some people said that its properties were better known than those of matter. Ether was the word most used by the theoretical physicist from Göttingen Professor Woldemar Voigt in his lectures on optics. And in summer 1905 all crucial experiments to detect the ether were giving negative results.[16] This seminar was the starting-point for Minkowski's celebrated work on electrodynamics of moving bodies.[17]

Born also recounted in 1955 in his talk "Physik und Relativität" ("Physics and Relativity") his vague impressions of the seminar,

"My first encounter with the difficulties, struggles which occurred in this orthodoxy in 1905, was in a seminar on the electron theory, led not by a physicist but by a mathematician, *Hermann Minkowski*. My memory of these old days is of course a little blurred. But I am quite sure that in this seminar we all discussed what was known at this time about the electrodynamics and optics of moving systems. We studied the works by *Hertz*, *FitzGerald*, *Larmor*, *Lorentz*, *Poincaré*, and others. As to it also, however, we also got an insight of *Minkowski's* own ideas, which were first published two years later".[18]

---

[12] Born, Max (1975/1978), Mein Leben: *Die Erinnerungen des Nobelpreisträgers*, 1975, München: Nymphenburger Verlagshandlung GmbH; *My Life Recollections of a Nobel Laureate*, 1978, New York, Charles Scribner's Sons, p. 80.
[13] Born, 1975/1978, pp. 82-83.
[14] Born, 1975/1978, p. 89.
[15] Born,1975/1978, p. 80.
[16] Born,1975/1978, p. 98.
[17] Born,1975/1978, p. 98.

Minkowski gave a lecture about Space and Time, under the title: "The Relativity Principle" on November 5, 1907 in the Göttingen Mathematical society. One month before this talk, on October 9, 1907, Minkowski had written to Einstein asking for a reprint of his 1905 paper, in order to discuss it in his seminar with Hilbert on the electrodynamics of moving bodies,[19]

"Dear Doctor Einstein,

At our seminar in the W.S. we also wish to discuss your interesting papers on electrodynamics. If you still have available reprints of your article in the *Ann. d. Phys. u. Ch.*, Vol. 17, I would be grateful if you would send us a copy. I was in Zürich recently and was pleased to hear from different quarters about the great interest being shown in your scientific successes.

With best regards, yours sincerely,

H. Minlowski".

In 1915, six years after Minkowski's death, Minkowski's 1907 talk was published in the *Annalen der Physik*, and coordinated to the *Annalen* by Arnold Sommerfeld.[20] Sommerfeld edited Minkowski's text, and introduced a few changes to the original manuscript. Sommerfeld expressed opposition to Minkowski's use of the ether; "Sommerfeld was unable to resist rewriting Minkowski's judgment of Einstein's formulation of the principle of relativity. He introduced a clause inappropriately praising Einstein for having used the Michelson experiment to demonstrate that the concept of absolute space did not express a property of phenomena. Sommerfeld also suppressed Minkowski's conclusion, where Einstein was portrayed as the clarifier, but by no means as the principal expositor, of the principle of relativity. With the exception of these changes, it seems that Sommerfeld printed Minkowski's manuscript substantially as he found it".[21]

On December 21, 1907, Minkowski talked to the Göttingen scientific society. On 5 April, 1908, his talk was published in *Göttinger Nachrichten* as a technical paper, "Die Grundgleichungen für die elektromagnetischen Vorgänge in bewegten Körpern"("The Basic Equations for Electromagnetic Processes in Moving Bodies").[22]

---

[18] Born Max, *Physics in My Generation: A selection of Papers*, 1969, London: Pergamon Press, p. 101; Born, Max, *Physik Im Wandel Meiner Zeit*, 1957, Germany: F. Vieweg, p. 186.
[19] Minkowski to Einstein, October 9, 1907, *CPAE*, Vol. 5, Doc. 62.
[20] Minkowski, Hermann (1907/1915), "Das Relativitätsprinzip" (Presented in Göttingen on 5.11.1907, Published Post-mortem by Arnold Sommerfeld), *Annalen der Physik* 47, 1915, pp. 927-938, p. 927.
[21] Pyenson, Lewis, *The Young Einstein: The Advent of Relativity*, 1985, Boston: Adam Hilger, p. 82.
[22] Minkowski, Hermann (1908a), "Die Grundgleichungen für die elektromagnetischen Vorgänge in bewegten Körpern, *Nachrichten von der Königlichen Gesellschaft der Wissenschaften zu Göttingen*, 1908, pp. 53-111.

This was Minkowski's only publication on the topic of electrodynamics and relativity to appear before his death on January 12, 1909. Pais writes that in this paper Minkowski for the first time presented the Maxwell-Lorentz equations in their modern tensor form, and the inadequacy of the Newtonian gravitational theory from the relativistic point of view is discussed.[23]

On September 21, 1908, in the 80th annual general meeting of the German Society of Scientists and Physicians (Gesellschaft Deutscher Naturforscher und Ärzte) at Cologne (Köln) in 1908, Minkowski presented his famous talk, "Raum und Zeit"; a culmination of Minkowski's burst of creativity of 1907-1908.

Minkowski's assistant, Max Born recounted,[24]

"One day [Fritz] Reiche asked me whether I knew a paper by a man named Einstein on the principle of relativity. He said Planck considered it most important. I had not heard of it, but when I learned that it had something to do with the fundamental principles of electrodynamics and optics which years ago had fascinated me in Hilbert and Minkowski's seminar, I agreed at once to join Reiche in studying it. From that moment relativity became our principal interest, and both [Stanislaw] Loria and [Rudolf] Ladenburg were infected by our enthusiasm. I began to think and work on some relativistic problems, and when I got stuck over some difficulties I wrote a letter to Minkowski asking his advice.

His reply was a great surprise. He did not answer my questions but said that he was himself working on the same subject and would like to have a young collaborator who knew something of physics, and of optics in particular. Would I like to come to Göttingen once again? And he added that this might lead to my future entrance into an academic career. Then he proposed that I should attend the annual meeting of the German Society of Scientists and Physicians, which was to be held in Cologne in September. There he could answer my questions and substantiate his proposition.

[…] I went to Cologne, met Minkowski and heard his celebrated lecture 'Space and Time', delivered on 21 September 1908. […] He told me later that it came to him as a great shock when Einstein published his paper in which the equivalence of the different local times of observers moving relative to each other was pronounced; for he had reached the same conclusions independently but did not publish them because he wished first to work out the mathematical structure in all its splendor. He never made a priority claim and always gave Einstein his full share in the great discovery.

After having heard Minkowski speak about his ideas, my mind was made up at once, I would go to Göttingen to help him in his work".

Scott Walter writes, "This story of Minkowski's recollection of his encounter with Einstein's paper on relativity is curious, in that the idea of the observable equivalence of clocks in uniform motion had been broached by Poincaré in one of the papers studied during the first session of the electron-theory seminar. It is possible, of course, that Poincaré's operational definition of local time escaped Minkowski's attention, or that Minkowski was thinking of an exact equivalence of timekeepers".[25]

---

[23] Pais, 1982, pp. 151-152.
[24] Born, 1975/1978, pp. 130-131.

Before 1905 Poincaré stressed the importance of the method of clocks and their synchronization by light signals. He gave a physical interpretation of Lorentz's local time in terms of clock synchronization by light signals, and formulated a principle of relativity. John Stachel explains: "Poincaré had interpreted the local time as that given by clocks at rest in a frame moving through the ether when synchronized as if – contrary to the basic assumptions of Newtonian kinematics – the speed of light were the same in all inertial frames. Einstein dropped the ether and the 'as if': one simply synchronized clocks by the Poincaré convention in each inertial frame and accepted that the speed of light really is the same in all inertial frames when measured with clocks so synchronized".[26]

In the Cologne talk Minkowski said immediately after presenting Lorentz's local time, "However, the credit of first recognizing sharply that the time of the one electron is just as good as that of the other, i.e., that $t$ and $t'$ are to be treated the same, is of A. Einstein". And Minkowski referred to Einstein's 1905 relativity paper and to his 1907 review article.[27]

Until 1912 Einstein's intuition was not strong enough in the field of mathematics, he considered mathematical formalism superfluous learnedness. Louis De Broglie distinguished between mathematical physics and theoretical physics.[28] The first, according to De Broglie, is the profound and critical examination of the physical theories put forward by the researcher who assesses mathematical speculations in order to improve these theories and in order to render their inherent proofs more rigorous. In contrast, theoretical physics is the construction of theories suitable to serve as an explanation of the experimental facts and to guide the work of the laboratory staff. Extensive mathematical knowledge is a pre-requisite, although it is not, ordinarily, the work of real mathematicians; it requires wide knowledge of the experimental facts, and mainly some kind of intuition in physics, which not all mathematicians have, as did Poincaré. Poincaré, according to De Broglie, was especially destined to engage fruitfully in mathematical physics. De Broglie asserted that Einstein was a scientist of the second type, and therefore he had succeeded in

---

arriving at the theory of relativity. In respect to Poincaré, "mathematical physics" was even apparent in the manner in which he presented his scientific endeavors.

The Cologne talk was Minkowski's last talk before he died. The talk had aroused great interest at the Cologne Congress and has become the basis of the modern mathematical apparatus of the relativity theory. Minkowski sent his manuscript a few days before Christmas – on December, 23, 1908 – to the editor of the *Physikalisch Zeitschrift*. Four days before his death the editor of the *Physikalisch Zeitschrift* spoke with Minkowski about the contents of his talk. He did not say what Minkowski told him, but mentioned that no one imagined that he would die suddenly. The *Physikalisch Zeitschrift* published the manuscript of the talk on 1909. [29]

In 1915 Minkowski's talk was reproduced in the book *Das Relativitätsprinzip*, a collection of the path breaking papers of special and one general relativity paper (with comments by Arnold Sommerfeld). The book began with an abridged version of Lorentz's 1904 paper on the theory of electron; then afterwards, were brought Einstein's 1905 two relativity papers. Minkowski's 1908 talk was reproduced in this book with minor changes.[30] Sommerfeld (as said above) also brought Minkowski's 1907 talk for publication in 1915.

The differences between the Cologne 1908 talk as published in 1909 and the same talk as it appeared in the 1915 book are quite critical in some places. For instance, at the end of the 1915 version we find the following sentence,[31]

"The validity without exception of the World postulate, so I would like to believe, is the true core of an electromagnetic worldview, which started by Lorentz, and further revealed by Einstein, in full day light" ["Die ausnahmslose Gültigkeit des Weltpostulates ist, so möchte ich glauben, der wahre Kern eines elektromagnetischen Weltbildes, der von Lorentz getroffen, von Einstein weiter herausgeschält, nachgerade vollends am Tage liegt"].

On the other hand, at the end of the manuscript of the 1908 talk as it was published in the *Physikalisch Zeitschrift* in 1909 we find,[32]

"The validity without exception of the Relativity postulate, so I would like to believe, is the true core of an electromagnetic worldview, which started by Lorentz, and further revealed by Einstein, in full day light" ["Die ausnahmslose Gültigkeit des Relativitätspostulates ist, so möchte ich glauben, der wahre Kern eines elektromagnetischen Weltbildes, der von Lorentz getroffen, von Einstein weiter herausgeschält, nachgerade vollends am Tage liegt"].

---

[29] Minkowski, 1908b, p. 104.
[30] Lorentz, Hendryk, Antoon., Einstein, Albert, and Minkowski, Hermann, *Das Relativitätsprinzip*, 1915, Leipzig und Berlin: Druck und Verlag von B. G. Teubner, pp. 56-73.
[31] Lorentz, Einstein, and Minkowski, 1915, p. 68.
[32] Minkowski, 1908b, p. 111.

The 1915 book was updated in 1922 [33] and translated to English in 1923 as *The Principle of Relativity*.[34] Minkowski's 1915 version above was translated from German to English and the above 1915 paragraph appeared in the 1923 English translation.

Minkowski never spoke about "The validity without exception of the World postulate", and this was *an editing* of his 1908 manuscript.

Minkowski started his 1908 talk with the famous paragraph,[35]

"M. H.! [ladies and gentleman!] The views of space and time, which I would like develop, have sprung from the experimental-physical soil. Therein lies their strength. They tend to be radical. Henceforth space by itself and time by itself, fade away completely into shadow, and only a kind of union of the two will preserve independent permanency".

And Minkowski presented new notions, "Weltpunkt x, y, z, t", "Kerwe in der Welt", "Weltlinie", and "Weltlinien".[36]

Minkowski considered a world, space-time; he examined the transformations that leave the expression:

$$c^2t^2 - x^2 - y^2 - z^2 = 1$$

Invariant. This expression had already appeared in his two previous papers. Minkowski said that the structure of this equation with positive $c$ consists of two hyperboloid sheets separated by $t = 0$. When transforming $x, y, z, t$ into $x', y', z', y', t'$, while $y$ and $z$ remain unchanged, we draw the upper branch of the hyperbola, the upper "light cone":

$$c^2t^2 - x^2 = 1.$$

Minkowski explained that when $c = \infty$ (c is the velocity of light in empty space), the group of Newtonian mechanics $G_\infty$ is graphically out of the hyperbola (the upper light cone). Hence the group $G_c$ is mathematically more intelligible than the Newtonian group $G_\infty$.[37]

In 1921 Max Born explained Minkowski's "Absolute World",[38]

---

[33] Lorentz, Hendryk, Antoon., Einstein, Albert, and Minkowski, Hermann, *Das Relativitätsprinzip*, 1922, Leipzig und Berlin: Druck und Verlag von B. G. Teubner. (Einstein's personal library, the Einstein Archives). This version already includes Einstein 1916 review paper on the General Theory of Relativity.
[34] Lorentz, Hendryk, Antoon, Einstein, Albert, and Minkowski, Hermann, *The Principle of Relativity*, 1923/1952, New York: Dover.
[35] Minkowski, 1908b, p. 104.
[36] Minkowski, 1908b, p. 104.
[37] Minkowski, 1908b, p. 105.
[38] Born, Max, *Die Relativitätstheorie Einsteins*, 1922, Berlin: Verlag von Julius Springer, pp. 218-219; Born Max, *Einstein's Theory of Relativity*, 1924/1962, New-York: Dover, p. 305.

"We have made use of his method of description throughout, by omitting the *y*- and *z*-axis only for the sake of simplicity and working in the *xt* plane. If we take a look at the geometry in the *xt*-plane from the mathematical point of view, we see that we are not dealing with ordinary Euclidean geometry. Since therefore, all straight lines that originate from the zero-point are equivalent, the unit of length on them is the same as the calibration curve, that is, a circle. But in the *xt*-plane the spacelike and timelike straight lines are not equivalent, there is a different unit of length on each, and the calibration curve consists of the hyperbole

$G = x^2 - c^2t^2 = \pm 1$."

An important property of Minkowski's world is the following: consider the *x-t* axes. Suppose we have a world-line, straight and parallel to the axis of *t*. This line corresponds to a point at rest. If the world-line is straight at an angle to the axis of *t*, this corresponds to a point in uniform motion. If the point is in non-uniform motion, the world line is a curved. We can draw new axes t'-x' in such a way that a substance that is in uniform motion in *x-t* is at rest in *x'-t'*. From $c^2t^2 - x^2 - y^2 - z^2 = 1$ Minkowski arrived at: $c^2dt^2 - dx^2 - dy^2 - dz^2$, which is positive, and thus the velocity *v* is always less than *c*.[39]

Minkowski explained that according to *Lorentz*, any moving body, experiences a contraction in the direction of motion with a velocity *v* in the ratio:

*1:* $\sqrt{(1 - v^2/c^2)}$.[40]

According to Einstein, the Lorentz contraction cannot be detected by physical means for an observer moving with the object. However, it can be detected by physical means by a non-co-moving observer. This is embodied in Minkowski's graphical space-time "world".

Minkowski presented his talk on September 12[th], 1908, and this talk had aroused great interest at the Cologne [Köln] Congress and has become the basis of the modern mathematical apparatus of the relativity theory. Einstein was supposed to meet Planck at the Cologne Congress, but Einstein did not come to this meeting. Planck wrote Einstein on September 9, 1908, "would it not be more practical if we were to meet in Cologne at the Naturforscherversammlung? I am certain to be there. There we would have more time and be in a more appropriate mood for scientific discussions".[41]

---

"Wir haben uns seiner Darstellung durchweg bedient, wobei wir nur zur Vereinfachung die *y*- und *z*-Achsen fortließen und in der *xt*-Ebene operierten. Werfen wir nun noch einen Blick vom mathematischen Standpunkte auf die Geometrie in der *xt*-Ebene, so erkennen wir, daß es sich nicht um die gewöhnliche Euklidische Geometrie handelt. Denn bei dieser sind alle vom Null punkt ausgehenden Geraden gleichberechtigt, die Längeneinheit auf ihnen ist dieselbe, die Eichkurve also ein Kreis. In der *xt*-Ehene aber sind die raumartigen und zeitartigen Geraden nicht gleichwertig, auf jeder gilt eine andere Längeneinheit, die Eichkurve besteht aus den Hyperbeln
$G = x^2 - c^2t^2 = \pm 1$."
[39] Minkowski, 1908b, p. 106.
[40] Minkowski, 1908b, p. 106.
[41] Planck to Einstein, September 9, 1908, *CPAE*, Vol. 5, Doc. 118.

Born arrived at the second time in Göttingen just a few weeks before Minkowski died,[42]

"I arrived in Göttingen rather late, on December 1908, long after the semester had started; […] The following day I called on Minkowski and was received very kindly. He explained to me his ideas on electrodynamics and relativity, and he listened patiently to my own suggestions. Once again I was fortunate in meeting a leading man in my branch of science and in being allowed to watch him working on a subject which fascinated me. But alas, it lasted only a few weeks. When I returned from a short Christmas holiday at home, I learned that Minkowski had been taken to hospital dangerously ill, and that an appendicitis operation had been performed. I did not see him again. He died on 12 January 1909. […]

Minkowski's numerous papers and unfinished manuscripts were given to [Andreas] Speiser, a young Swiss mathematician, and myself for sifting and possibly finishing.[43] We first of all separated the pure mathematics from mathematical physics; the latter was entrusted to me. It was formidable bundle, but most of it turned out to be either manuscripts of papers already published, or indecipherable sketches. Only one investigation which I knew in outline from my last interview with Minkowski was far enough advanced for me to try to finish it for publication. This I succeeded in doing. The article appeared in May 1910, together with a reprint of Minkowski's celebrated paper "Die Grundgleichungen für die Elektromagnetischen Vorgänge in Bewegten Körpern" (The basic equations for electromagnetic processes in moving bodies) as No 1 of a series of monographs edited by Blumenthal".[44]

*I wish to thank Prof. John Stachel from the Center for Einstein Studies in Boston University for sitting with me for many hours discussing special relativity and its history.*---

[42] Born, 1975/1978, pp. 132-133.
[43] Minkowski, Hermann, *Gesammelte Abhandlungen von Minkowski*, Hermann unter mitwirkung von Andreas Speiser und Hermann Weyl herausgegeben David Hibert, Zweiter band, mit einem bildnis Hermann Minkowskis 34 Figuren im text und einer doppeltafel, 1911, Leipzig: G. B. Teubner.
[44] Minkowski, Hermann, *Zwei Abhandlungen über die Grundgleichungen der Elektrodynamik*, Mit einem einführungswort von Otto Blumenthal, 1910, Leipzig und Berlin: B. G. Teubner.